\documentclass[epj]{svjour}

\def\be{\begin{equation}} 
\def\ee{\end{equation}}   
\RequirePackage{graphicx} 

\usepackage{graphics}

\usepackage{cite} 
\usepackage{amsmath}
\usepackage{hyperref}
\usepackage{cite}
\usepackage{amsmath,amssymb}
\usepackage{xcolor}

\begin{document}
\title{The effects of running gravitational coupling on three dimensional black holes}
\author{
\'Angel Rinc\'on \inst{1} 
\thanks{E-mail: \href{mailto:angel.rincon.r@usach.cl}{\nolinkurl{angel.rincon.r@usach.cl}} }
\and 
Benjamin Koch \inst{2}\; \inst{3}\; \inst{5}
\thanks{E-mail: \href{mailto:benjamin.koch@tuwien.ac.at}{\nolinkurl{benjamin.koch@tuwien.ac.at}} }
\and 
Cristobal Laporte \inst{4}
\thanks{E-mail: \href{mailto:cristobal.laportemunoz@ru.nl}{\nolinkurl{cristobal.laportemunoz@ru.nl}} }
\and 
Felipe Canales \inst{5}
\thanks{E-mail: \href{mailto:facanales@uc.cl}{\nolinkurl{facanales@uc.cl}} }
\and 
Norman Cruz \inst{1} \; \inst{6} 
\thanks{E-mail: \href{mailto:norman.cruz@usach.cl}{\nolinkurl{norman.cruz@usach.cl}} }
}                     
%
%
\institute{
Departamento de Física, Universidad de Santiago de Chile, Avenida Ecuador 3493, Santiago, Chile
\and
Institut fur Theoretische Physik, Technische Universitat Wien, Wiedner Hauptstrasse 8-10, A-1040 Vienna, Austria 
\and
Atominstitut, Technische Universitat Wien, 
Stadionalle 2, A-1020 Vienna, Austria 
\and
Institute for Mathematics, Astrophysics and Particle Physics (IMAPP), Radboud University, Heyendaalseweg 135, 6525 AJ Nijmegen, The Netherlands 
\and
Instituto de F{\'i}sica, Pontificia Universidad Cat{\'o}lica de Chile, Av. Vicu{\~n}a Mackenna 4860, Santiago, Chile 
\and
Center for Interdisciplinary Research in Astrophysics and Space Exploration (CIRAS), Universidad de Santiago de Chile, Av. Libertador Bernardo O'Higgins 3363, Estación Central, Chile.
}
\date{Received: date / Revised version: date}
%
\abstract{
In the present work, we investigate the consequences of
running gravitational coupling on the properties of the 
three-dimensional BTZ black hole. We take as starting point the functional form of gravitational coupling
obtained in the context of asymptotic safe gravity theory. By using the standard scale setting relation where $k\sim \xi/r^n$,
we compute the solution of the Einstein field equations. We get and analyze the horizon and the thermodynamic properties of this new class of black hole solutions. The impact of the scale--dependent parameter $\xi$ on the cosmological ``constant" and metric functions are briefly discussed. We find that the null energy condition is also violated in this setup when scale-dependent gravity and Newton's coupling (coming from the asymptotic safety scenario of gravity) are simultaneously taken into account. 
%
%
\PACS{
      {PACS-key}{discribing text of that key}   \and
      {PACS-key}{discribing text of that key}
     } 
} 
\maketitle

\tableofcontents

\section{Introduction} \label{intro}

A well--defined theory of quantum gravity remains as an open problem and it is still a challenge in theoretical physics. Up to now, at least 16 major approaches to quantum gravity have been proposed in the literature (see \cite{Esposito:2011rx} and references therein), but none of these approaches has completely solved the problem in a closed way. 
The most popular approaches are probably
loop quantum gravity (LQG) \cite{Rovelli:1997yv,Chiou:2014jwa}, certain modified gravity theories \cite{Clifton:2011jh,Koyama:2015vza}, as well as string theory \cite{Polchinski:1998rq,Polchinski:1998rr,Dienes:1996du}.  The aforementioned techniques are able to get insights into certain cosmological/astrophysical problems, e.g., Dark Energy, Dark Matter, as well as the physics of black holes and singularities of General Relativity. 
Even though these  approaches are very different, they have the common feature
that they can be mapped to  effective
generalizations of GR.

In this context of theories beyond Einstein gravity, we should mention a particularly well known generalization: the Brans-Dicke theory of gravity \cite{Brans:1961sx,Brans:1962zz}. 
In Brans-Dicke (BD) theory, the strength of the gravitational interaction is controlled by a scalar field (Newton's coupling). 
Thus, Newton's coupling is identified as a scalar field, $G \rightarrow \phi^{-1}$. 
This leads to a characteristic contribution to the classical field equations, which accounts for the modification of Newton's coupling. 
Based on this approach, there are many examples where the basic parameters that enter into the effective action defining the model (Newton’s coupling, the electromagnetic coupling, and the cosmological constant, among others) become scale-dependent (SD) functions
\cite{Sola:2015wwa,Sola:2017znb,Torres:2017ygl,Ishibashi:2021kmf,Sendra:2018vux,Saueressig:2015xua,Koch:2014cqa,Falls:2012nd,Koch:2013rwa,Bonanno:2001xi,Bonanno:2006eu,Reuter:2006zq}
. 
Such an effect is quite common since scale-dependence (at the level of the effective action) is a generic feature of effective quantum field theory. 

The inclusion of quantum features can be systematically accounted for at different levels. Roughly speaking, we can identify (at least) three different ways to modify the classical solutions. One can categorize them as follows~\cite{Reuter:2003ca}:
\begin{itemize}
    \item[i)] at the level of solution, 
    \item[ii)] at the level of the equations of motion, and
    \item[iii)] at the level of the action.
\end{itemize}

While the first modification is by construction
only reasonable for tiny modifications of classical solutions, the other two modifications can, in principle, give reasonable results beyond this limit. 
In each of these applications, it is of crucial importance for the observables how the quantum scale-dependence
and the corresponding scale-setting is implemented.
Several conceptually different approaches that can be implemented in
one of these modifications. Some of these approaches are
\begin{itemize}
    \item[a)] use the beta functions of the gravitational theory (e.g. Asymptotic safety)
    to derive the SD couplings $G_k, \dots$ and set the scale in terms of a physical quantity describing the observed system.
    \item[b)] treat the couplings, in particular, the gravitational coupling as independent field $G=G(x)$ and impose a dynamic nature to this field, for example, by introducing a kinetic term for this field \cite{Brans:1961sx,Brans:1962zz}.
    \item[c)] impose a scaling nature for $G(x)$ which is motivated by the approach $a)$ and then 
    solve the corresponding field equations from $ii)$. This approach is implemented in this paper.
\end{itemize}
In theories of gravity, the scale-dependence is expected to modify the horizon, the thermodynamics as well as the quasi normal spectra of classical black hole backgrounds. Also, other well-known black hole effects have been reviewed in light of this formalism. To be more precise, we can mention: i) the Sagnac effect \cite{Rincon:2019zxk} and, ii) the evolution of trajectories of photons \cite{Fathi:2019jid}. In the cosmological context there are a few novel solutions within this formalism. The same occurs in the level of wormhole solutions and relativistic compact stars. For an exhaustive history of the implementation of the formalism above-mentioned, please see
\cite{Koch:2016uso,Rincon:2017ypd,Rincon:2017goj,Rincon:2017ayr,Contreras:2017eza,Rincon:2018sgd,Contreras:2018dhs,Rincon:2018lyd,Rincon:2018dsq,Contreras:2018gct,Canales:2018tbn,Rincon:2019cix,Panotopoulos:2021heb,Bargueno:2021nuc,Contreras:2018swc,Koch:2015nva,Contreras:2013hua,Alvarez:2020xmk,Rincon:2020cpz,Rincon:2020iwy} and references therein.

In this paper, we make some progress on the topic of quantum
gravity by studying the well-known BTZ black hole assuming an effective scale--dependent gravity in 2+1 dimensions with a cosmological constant. With this in mind, we then mix three different aspects, i.e., scale dependence, black holes and gravity in three dimensions. They are, by themselves, a good inspiration to go beyond GR because those aspects have an important motivation from the perspective of quantum gravity.

This paper is organized as follows: after this introduction, we review the classical BTZ black hole solution without angular momentum in  Sect.~\eqref{sec:1}. Subsequently, in Sect.~\eqref{sclike} we show the basic ingredients of the SD like solutions and their properties, i.e.,
the potential relation between SD gravity and asymptotically safe gravity, 
the concrete solutions, 
the horizon and invariants, 
the thermodynamics, and 
a comparative discussion of RG (RG) improvement and SD gravity in the context of black hole physics. 
Before concluding, we generalize the discussion by including a rotational degree of freedom.
Finally, in Sect.~\eqref{conclu} we summarize our main findings.

\section{Review of BTZ black hole solution} \label{sec:1}

This section is devoted to the classical BTZ black hole solution \cite{Banados:1992wn,Banados:1992gq},
its line element, event horizons, and thermodynamics.
At first, the contribution of angular momentum will be neglected. 
We will focus on the Einstein Hilbert action with a cosmological term, which is the mi\-nimal coupling between gravity and matter, and it is given by 
\be\label{actionBTZ}
I_0[g_{\mu \nu}] = \int {\mathrm {d}}^3x \sqrt{-g} \Bigg[\frac{1}{2 \kappa_0}\Bigl(R - 2\Lambda_0\Bigl) \ + \ \mathcal{L}_{M} \Bigg],
\ee
where $g$ is the determinant of the metric field, $\mathcal{L}_{M}$ is the matter Lagrangian, $\Lambda_0$ is the cosmological constant, 
$\kappa_0 \equiv 8 \pi G_0$ is the gravitational coupling, $R$ is the Ricci scalar, and finally $g_{\mu \nu}$ is the metric field. 
To obtain the Einstein field equations, we vary the classical action with respect to the metric field, 
leading to the field equations
\be\label{eomBTZ}
 G_{\mu\nu} + \Lambda_{0} g_{\mu\nu} = \kappa_0 T_{\mu \nu},
\ee
where $T_{\mu \nu}$ is the energy momentum tensor associated to a matter content defined as follow
\begin{align}
T_{\mu \nu} &\equiv T^{M}_{\mu \nu} = -2 \frac{\delta \mathcal{L}_{M}}{\delta g^{\mu \nu}} + \mathcal{L}_M g_{\mu \nu} .
\end{align}
Following the conventional route, in spherically symmetric spacetimes, the line element is parametrized as
\begin{align}\label{lineele}
ds^2= -A_0(r) dt^2+ B_0(r) dr^2 + r^2 d\phi^2,
\end{align}
and, solving the Einstein's field equation, one obtains the conventional vacuum, spherically symmetric, neutral and non-rotating BTZ black holes \cite{Banados:1992wn,Banados:1992gq}, namely
%
\begin{align}
A_0(r) &\equiv -M_0 + \left(\frac{r}{\ell_0}\right)^2,
\\
B_0(r) &\equiv A_0(r)^{-1}, 
\\
\Lambda_0 &\equiv - \frac{1}{\ell_0^2} .  
\end{align}
 By demanding that $A(r_0)=0$
one obtains the horizon radius
\begin{align}
 r_0 &= \ell_0 M_0^{1/2}  .
\end{align}
The curvature invariants of the solution are 
\begin{align}
    R_0 &\equiv -\frac{6}{\ell_0^2},
    \\
    K_0 &\equiv \frac{12}{\ell_0^4}.
\end{align}
The corresponding Hawking temperature and the Bekenstein-Hawking entropy are 
\begin{align}
T_0(r_0) &= \frac{1}{4 \pi} \Bigg| \frac{2 G_0 M_0}{r_0}  \Bigg|,
\\
S_0(r_0) &= \frac{\mathcal{A}_{H}}{4 G_0},
\end{align}
the black hole area $\mathcal{A}_H$ is then 
\begin{align}
\mathcal{A}_{H}(r_0) &= \oint  {\mathrm {d}}x \sqrt{h} = 2 \pi r_0.
\end{align}
Finally, the heat capacity is written as
\begin{align}
C_0(r_0) &= T \ \frac{\partial S}{\partial T} \ \Bigg{|}_{r_0} = S_0.
\end{align}

\section{Scale--dependent like solution} \label{sclike}

In the context of scale--dependent gravity, we can solve the gap equations of motion~\cite{Avan:1983bv} exactly and afterward obtain the metric functions and the corresponding shape of the running couplings. This is the usual approach followed in several previous works, but now we will move to a  closely related but different approach, where one takes advantage of the exact results provided by the asymptotic safety (AS) program.
%
%
The key ingredient for investigating the asymptotic safety (AS) scenario is the gravitational average effective action, a Wilson-type effective action that dictates the evolution of the SD~\cite{Wetterich:1992yh,Morris:1993qb,Reuter:1993kw} couplings,
\begin{equation}\label{Wetterich}
    \partial_t \Gamma_k = \frac{1}{2} \text{STr}\left[\left(\Gamma^{(2)}_k + \mathcal{R}_k\right)^{-1}\partial_t \, \mathcal{R}_k\right].
\end{equation}
The formulation of the functional renormalization group equation~(\ref{Wetterich}) relies on the introduction of the infrared regulator, while the supertrace represents a sum over all internal indices as well as an integration over spacetime. For the study of three-dimensional BTZ black holes, we restrict the gravitational part of $\Gamma_k$ to the euclidean Einstein-Hilbert (EH) action,
\begin{align}\label{eq_EH}
\Gamma^{\text{EH}}_{k}[g_{\mu \nu}] &= 
 \int {\mathrm {d}}^3x \sqrt{-g} \Bigg[\frac{1}{2 \, \kappa_k}\Bigl(R - 2\Lambda_k \Bigl) \ + \ \mathcal{L}_{M} \Bigg],
\end{align}
The effective field equations, obtained from a variation of \eqref{eq_EH} with respect to $g_{\mu \nu}(x)$, are \cite{Rincon:2017ayr}:
\begin{align}\label{eq_eomSD}
R_{\mu \nu} - \frac{1}{2}R g_{\mu \nu} + \Lambda(r) g_{\mu \nu} = - \Delta t_{\mu \nu}
\end{align}
where the G-varying part $\Delta t_{\mu \nu}$ is computed to be \cite{Rincon:2017ayr}
\begin{equation}
\Delta t_{\mu \nu} = G_k \Bigl(g_{\mu \nu} \Box - \nabla_\mu \nabla_\nu\Bigl) G_k^{-1}.
\end{equation}
Further, 
when the truncation (\ref{eq_EH}) (supplemented by a gauge-fixing and ghost actions) is inserted back in Eq.~(\ref{Wetterich}), the beta functions for the three-dimensional gravitational coupling and their corresponding anomalous dimension using a Litim-regulator in the Feynman-de Donder gauge reads \cite{Reuter:1996cp,Reuter:2001ag,Niedermaier:2006wt,Reuter:2012id}, 
\begin{align}
    \beta_g = \left(1 + \eta_N\right)\hspace{.1 cm},\hspace{.1 cm} \eta_N = \frac{B_1 \, g}{1 - g \, B_2}
\end{align}
where the coefficients $B_{1,2}$ depend on the cosmological constant as well as various technical details involved in the implementation of RG techniques. In the limit where $|B_1|\gg |B_2|$, the resulting $G(r)\equiv G(k = k(r))$ is approximately given by, 
\begin{align}\label{G(r)I}
    G(r) = \frac{G_0}{1 + G_0 \, B_1 \, k}.
\end{align}
Once one knows the solutions of the RG equations, the scale needs to chosen.
This can be done following different approaches \cite{Platania:2019kyx,Eichhorn:2021etc,Eichhorn:2021iwq,Held:2021vwd}. 
One possible pathway to preserve the diffeomorphism invariance of the equations is identifying the RG scale as a carefully chosen function of dynamical variables. 
This identification results from applying the variational principle to $k$ by promoting $k$ to a field at the level of the effective action and setting the RG scale in terms of dynamical variables \cite{Koch:2010nn,Domazet:2012tw,Koch:2014joa,Koch:2020baj}.
In this work we use a inverse power law scale-setting, $k \equiv \left(\zeta/r\right)^2$, where $\zeta$ is an undetermined constant. 
This type of straightforward power-law makes sure that 
quantum gravity corrections vanish for large radii.
Redefining the denominator of $G(r)$ as $\xi^2 \equiv G_0 \, B_1 \, \zeta^2$, Eq.~(\ref{G(r)I}) can be written as,
\begin{align}\label{eq_Gr}
G(r) &= G_0 \Bigg[1 + \left(\frac{\xi}{r}\right)^2 \Bigg]^{-1}.
\end{align}
In the rest of the paper, $G_0=1$.  
%
\subsection{Solution} \label{sec:2}
Taking advantage of the particular form of the gravitational coupling $G(r)$, we can
solve the field equations (\ref{eq_eomSD}). 
The metric functions $A(r)$, $B(r)$ as well as the cosmological parameter $\Lambda(r)$ 
are found to be
\begin{align}
\begin{split}\label{eq_Asol}
A(r)  &=  A_0(r) + \frac{1}{3}M_0 
\Bigg[
24 - 6 \left(\frac{\xi}{r}\right)^2 + \left(\frac{\xi}{r}\right)^4
\\
& \hspace{1.35cm} - 24 \left(\frac{\xi}{r}\right)^{-2} \ln \left( 1 + \left(\frac{\xi}{r}\right)^2\right)
\Bigg],
\end{split}
\\ \label{eq_Bsol}
B(r) &=  \Bigg[ 1 - \left(\frac{\xi}{r}\right)^2 \Bigg]^6   A(r)^{-1},
\\
\Lambda(r) &=  -   \frac{A(r) \left(1 + \frac{\xi ^2}{r^2}\right) \left(1 - 3\frac{ \xi ^2}{r^2}\right)+M_0 \left(1-\frac{\xi ^2}{r^2}\right)^4}{r^2 \left(1-\frac{\xi }{r}\right)^6 \left(1 + \frac{\xi }{r}\right)^6 \left( 1 + \frac{\xi ^2}{r^2}  \right)^2}.
\end{align}
At this point, some comments are in order. It should be noticed that the lapse function $A(r)$ can be split into the classical part, $A_0(r)$, plus corrections intrinsically related to the scale--dependent scenario and parametrized by the energy scale $k \propto (\xi/r)^2$. What is more, we observe an unusual property of the line element: the combination $g_{tt} g_{rr} \neq -1$ which is surprising since it differs from the findings
in numerous other results in the context of SD black holes
\cite{Rincon:2020iwy}
and the so--called SD scenario 
\cite{Koch:2016uso,Rincon:2017ypd,Rincon:2017goj,Rincon:2017ayr,Contreras:2017eza,Rincon:2018sgd,Contreras:2018dhs,Rincon:2018lyd,Rincon:2018dsq,Contreras:2018gct,Canales:2018tbn,Rincon:2019cix,Panotopoulos:2021heb,Bargueno:2021nuc,Contreras:2018swc,Koch:2015nva,Contreras:2013hua,Alvarez:2020xmk,Rincon:2020cpz,Rincon:2020iwy}.
In these studies it was also shown that if the so called Null Energy Condition (NEC) is fulfilled,
then necessarily, the metric potentials $g_{tt}\cdot g_{rr}=1$.
Thus, the NEC is violated by the solution (\ref{eq_Asol}, \ref{eq_Bsol}).
Nevertheless, this new solution still recovers the classical solution when we turn off the dimensionless parameter $\xi = \xi/x_0$. Thus we have:
\begin{align}
\lim_{\xi \rightarrow 0} A(r) &= A_0(r) ,
\\
\lim_{\xi \rightarrow 0} B(r) &= A_0(r)^{-1},
\\
\lim_{\xi \rightarrow 0} \Lambda(r) &= \Lambda_0,
\end{align}
Thus, the new solution contains the classical one and also modifies the black hole properties at short distances. 

\begin{figure*}[ht]
\centering
\includegraphics[width=0.48\textwidth]{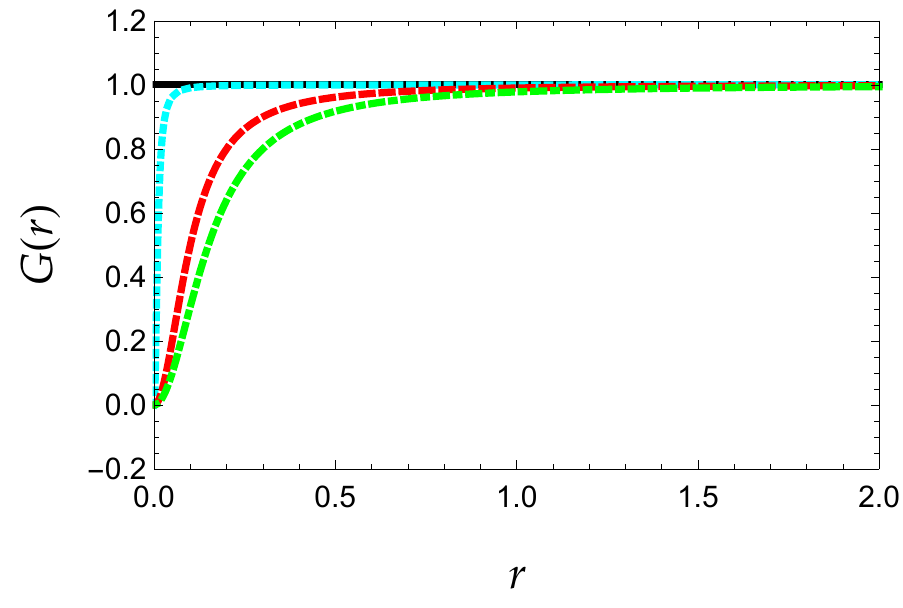}   
\ \ \
\includegraphics[width=0.48\textwidth]{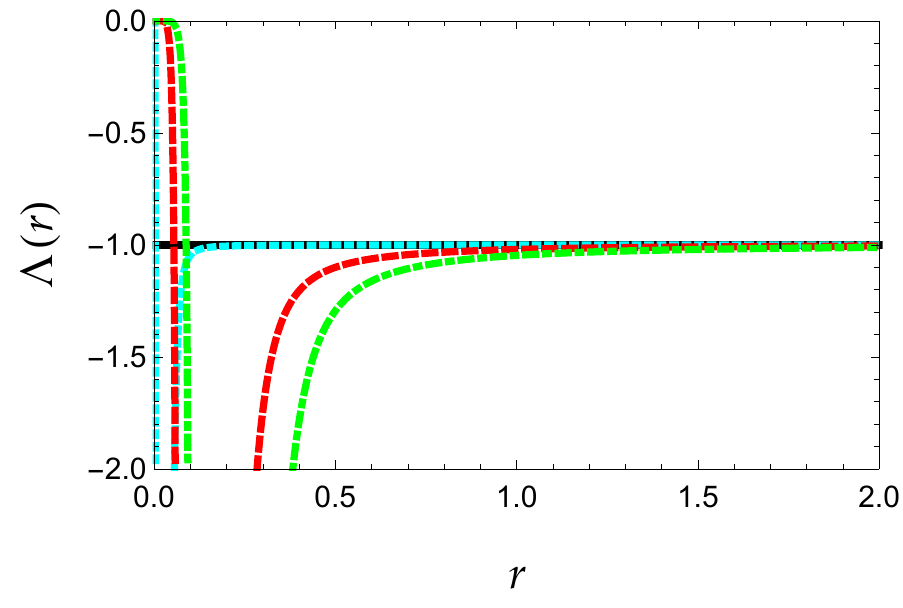}   
\caption{
{\bf{Left panel:}} Gravitational Newton's coupling from a given parametrization in asymptotically safe gravity.
{\bf{Right panel:}} Cosmological coupling $\Lambda(r)$ for a given $G(r)$.  In all four figures we fix $G_0 = 1, M_0 = 1$ and $\ell_0=1$. The color code is as follows:
  i) solid black line, for $\xi=0$,
 ii) dotted blue line, for $\xi=0.010$
iii) dashed red line, for $\xi=0.100$
iv) dot-dashed green line, for $\xi=0.150$
}
\label{fig:GandLambda}
\end{figure*}

\subsection{Horizon and invariants} \label{sec:2}

\begin{figure*}[ht]
\centering
\includegraphics[width=0.48\textwidth]{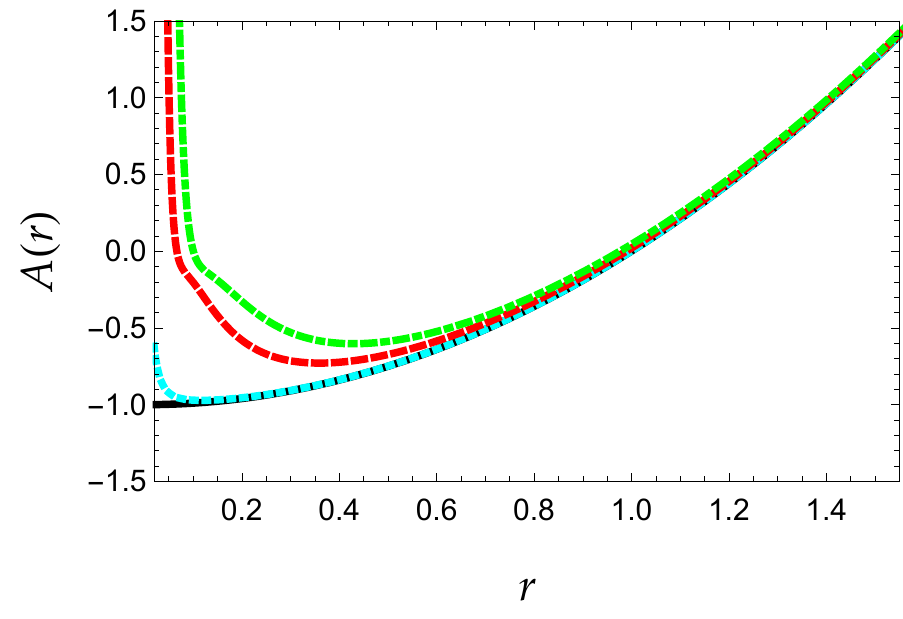}   
\ \ \
\includegraphics[width=0.48\textwidth]{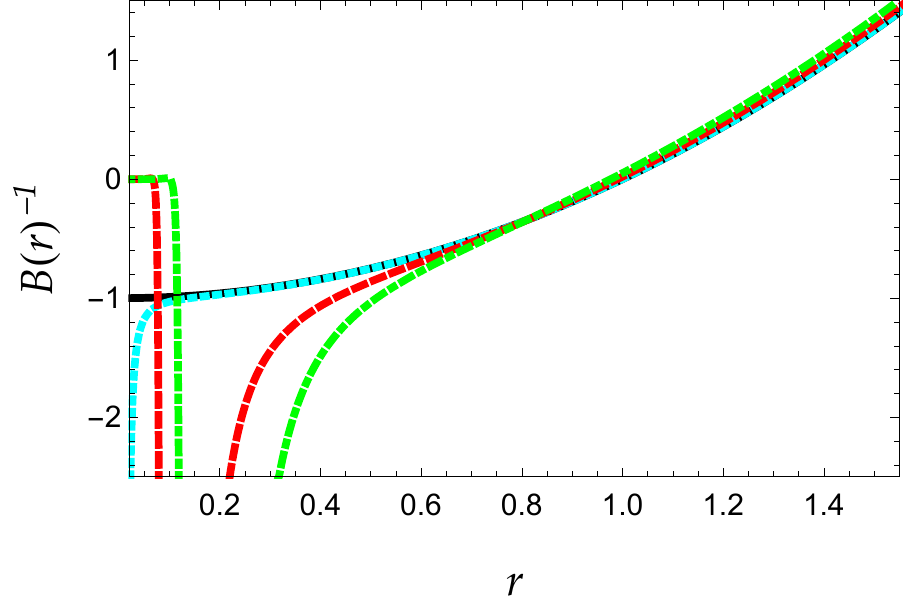}   
\caption{
Metric potentials for the obtained solutions for different values of $\xi$ assuming a concrete parametrization of Newton's coupling inspired in asymptotically safe gravity.
{\bf{Left panel:}} $A(r)$ versus radial coordinate.
{\bf{Right panel:}} $B(r)^{-1}$ versus radial coordinate. 
In all four figures we fix $G_0 = 1, M_0 = 1$ and $\ell_0=1$. The color code is as follows:
 i) solid black line, for $\xi=0$,
 ii) dotted blue line, for $\xi=0.010$
iii) dashed red line, for $\xi=0.100$
iv) dot-dashed green line, for $\xi=0.150$
}
\label{fig:potentials}
\end{figure*}

\begin{figure*}[ht]
\centering
\includegraphics[width=0.48\textwidth]{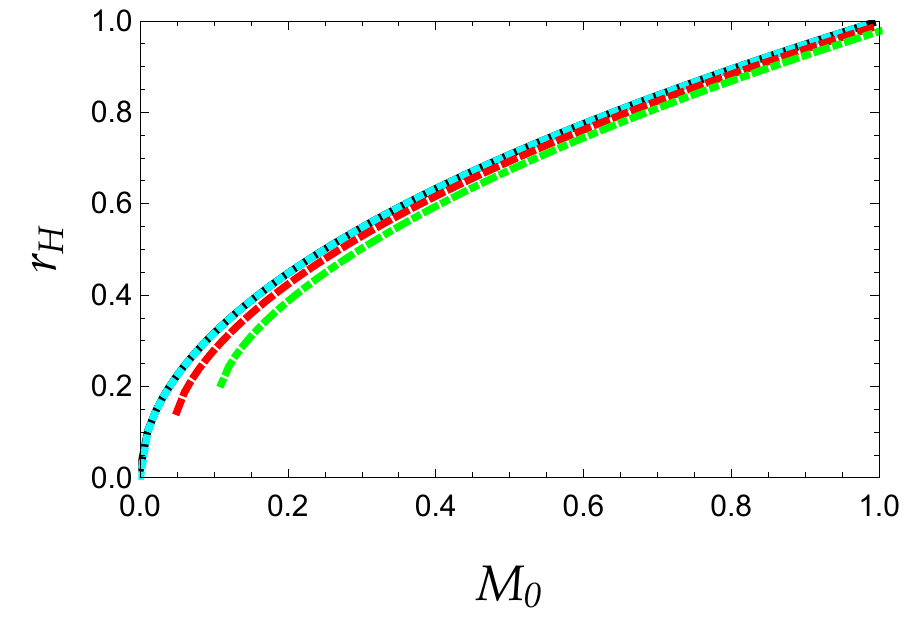}   
\ \ \
\includegraphics[width=0.48\textwidth]{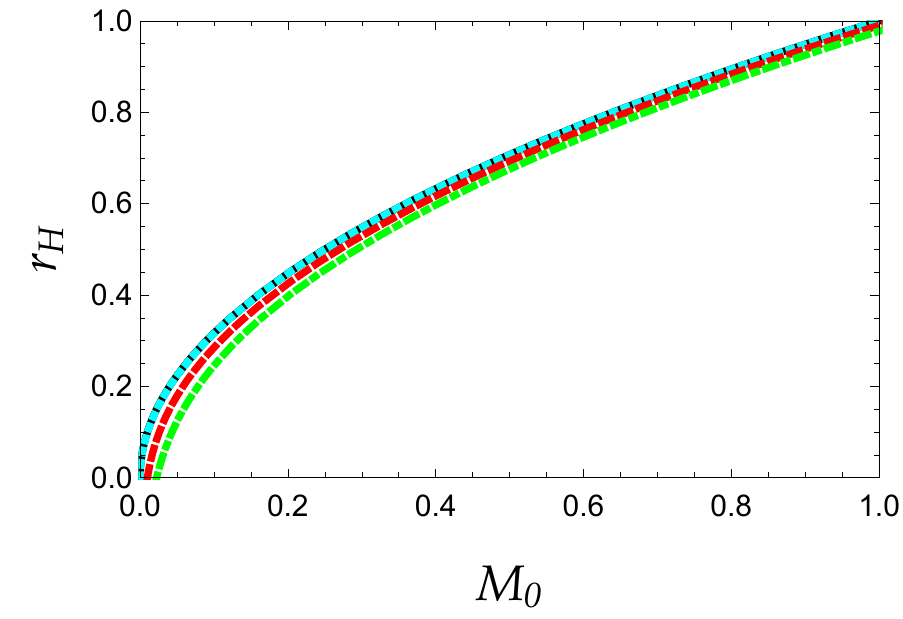}   
\caption{
Black hole horizon $r_H$ against the classical black hole mass $M_0$ for different values of the parameter $\xi$.
{\bf{Left panel:}} Exact black hole horizon obtained numerically by solving $B(r_H)^{-1}=0$, versus $M_0$. {\bf{Right panel:}} Approximated horizon obtained for a small value of the 
parameter $\xi$, maintaining just the leading correction.
The color code is as follows:
i) solid black line, for $\xi=0$,
 ii) dotted blue line, for $\xi=0.010$
iii) dashed red line, for $\xi=0.100$
iv) dot-dashed green line, for $\xi=0.150$
}
\label{fig:horizons}
\end{figure*}

At this point, we should investigate how the lapse function evolves. %
In particular, the event horizon, $r_H$, is obtained demanding $g^{rr}(r_H)=0$, which means that at $r = r_H$, the corresponding hypersurface is everywhere null. In the context of classical gravity, the BTZ solution has a simple and analytical expression for its horizon. However, this does not hold for the quantum counterpart. 
In particular, given the logarithmic contribution, no analytic solutions are found. Despite of that, one still can make progress for a certain region of the solution when $\xi$ is small enough. Thus, taking $g^{rr}(r_H)=0$ and considering $\xi$ small we have 
\begin{align}
-M_0 + \frac{r^2}{\ell_0^2} + \frac{ 6\xi^2}{\ell_0^2}-\frac{4 M_0 \xi^2}{r^2} =0,
\end{align} 
where the horizon is given by
\begin{align}
r_H^2 &= \frac{1}{2}r_0^2 \left[1 - \frac{6 \xi^2}{r_0^2} + \sqrt{1 + \frac{4\xi^2}{r_0^2} +  \frac{36\xi^4}{r_0^4}}  \right].
\end{align}
Please, notice that the value $r_0$ is the classical horizon provided by $B_0(r_0)^{-1}=0$. We also observe that when $\xi \rightarrow 0$, the classical BTZ solution is immediately recovered.
%
It is essential to point out that the black hole horizon is always smaller than the classical counterpart, as can be verified by making a subsequent approximation to the event horizon. Thus, when the parameter $\xi$ is small, we finally obtain
\begin{align} \label{raprox}
    r_H &\approx r_0\Bigg[1 - \left(\frac{\xi}{r_0}\right)^2\Bigg]
\end{align}
Note that an unexpected pole at $r \sim \xi$, absent in the classical solution, seems to emerge in this approach.
However, since (\ref{raprox}) is obtained from an expansion in
small $\xi$, this pathology lies outside the validity of the expansion.
The region with
$r \le \xi$, is not covered in this approximation and the black hole horizon is then computed in the region $r > \xi$. 
%
Fig.~\eqref{fig:potentials} shows the corresponding metric potentials. The left panel shows $A(r)$ against the radial coordinate, whereas the right panel shows $B(r)^{-1}$, again versus the radial coordinate.
In figure \eqref{fig:potentials} (right), notice: 
i)  the clear divergence when $r \rightarrow \xi$, and
ii) the horizon decreases when $\xi$ increases.
In this solution should be noticed the non-trivial relation between $A(r)$ and $B(r)$. Thus, when $\xi \rightarrow 0$, the Schwarzschild ansatz is retained; otherwise, such condition is violated. In this case, such violation is directly related to the approach used to obtain the explicit form of Newton's coupling.
Fig.~\eqref{fig:horizons} shows the black hole horizon in two different cases: 
i) the numerical computation, obtained by computing $B(r_H)^{-1}=0$ (left panel), and 
ii) the approximated computation, considering leading corrections on $\xi$ only (according to expression \eqref{raprox} on right panel).

\begin{figure*}[ht]
\centering
\includegraphics[width=0.48\textwidth]{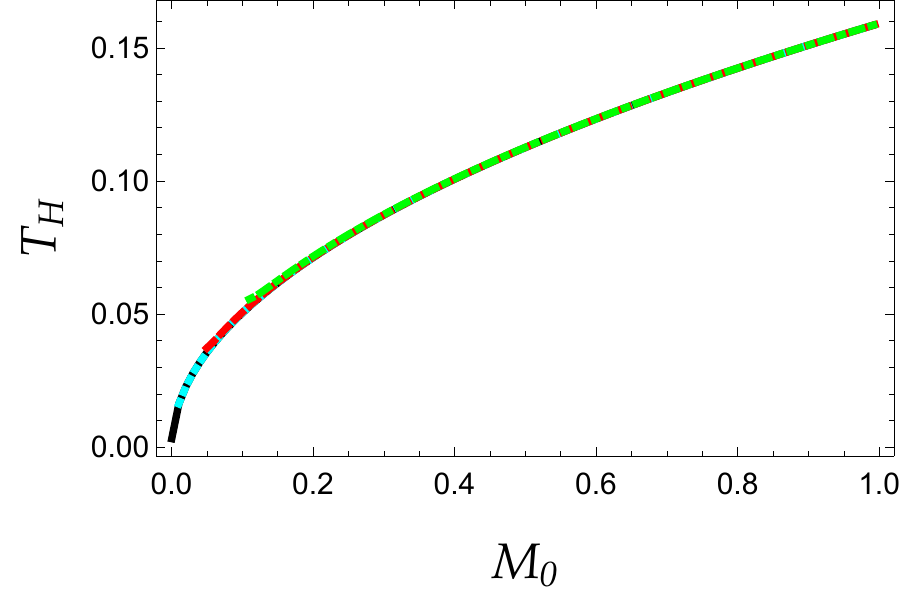}   
\ \ \
\includegraphics[width=0.48\textwidth]{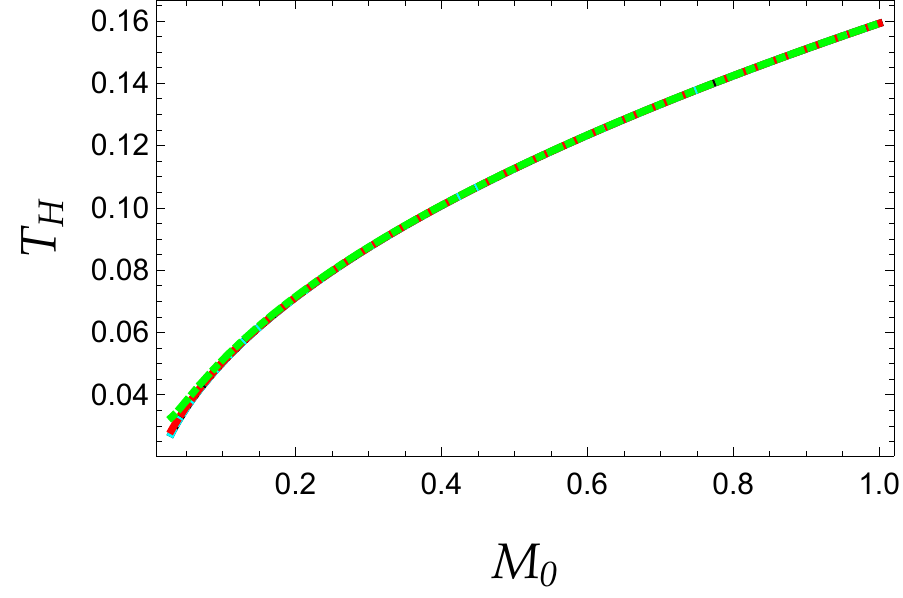}   
\caption{
Black hole temperature $T_H$ against the classical black hole mass $M_0$ for different values of the parameter $\xi$.
{\bf{Left panel:}} Exact black hole temperature obtained numerically by solving Eq.~\eqref{temp_exact}, versus $M_0$.
{\bf{Right panel:}} Approximated black hole temperature obtained for a small value of the 
parameter $\xi$, maintaining just the leading correction (using Eq.~\eqref{temp_approx}).
The color code is as follows:
  i) solid black line, for $\xi=0$,
 ii) dotted blue line, for $\xi=0.010$
iii) dashed red line, for $\xi=0.100$
iv) dot-dashed green line, for $\xi=0.150$.
}
\label{fig:temperature}
\end{figure*}

\begin{figure*}[ht]
\centering
\includegraphics[width=0.48\textwidth]{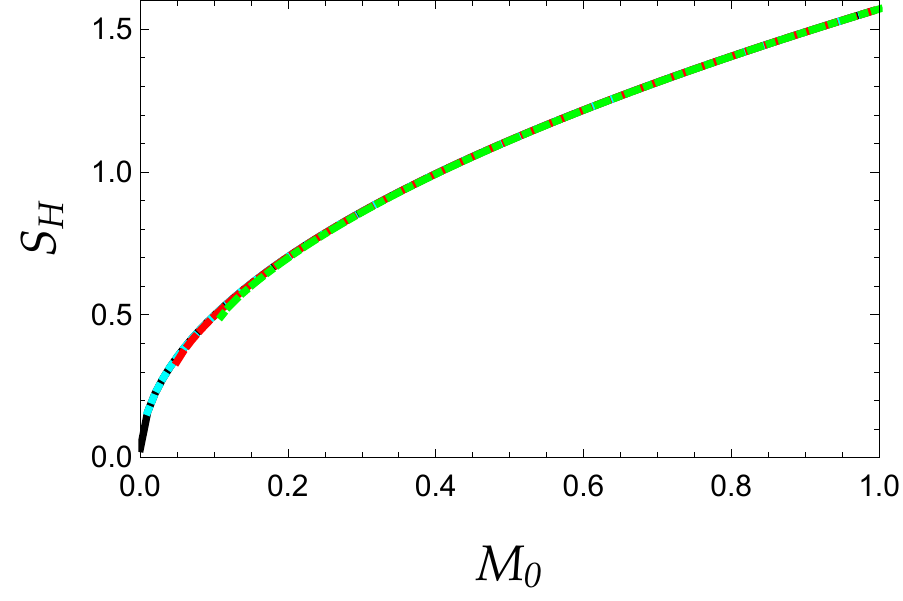}   
\ \ \
\includegraphics[width=0.48\textwidth]{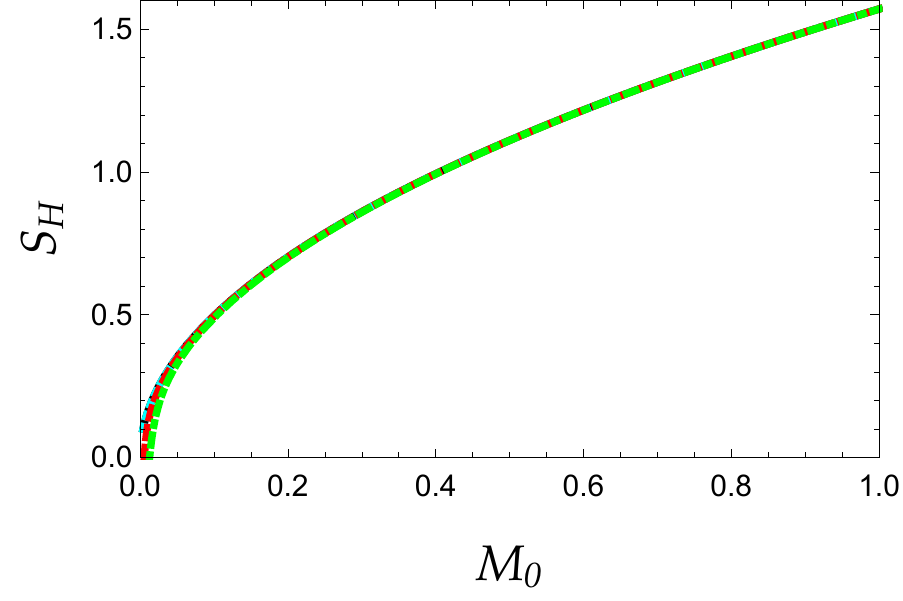}   
\caption{
Bekenstein Hawking entropy $S_H$ against the classical black hole mass $M_0$ for different values of the parameter $\xi$.
{\bf{Left panel:}} Exact black hole entropy obtained numerically by solving Eq.~\eqref{Entropy}, versus $M_0$.
{\bf{Right panel:}} Approximated black hole entropy obtained for a small value of the 
parameter $\xi$, maintaining just the leading correction (using Eq.~\eqref{Entropy2}).
The color code is as follows:
  i) solid black line, for $\xi=0$,
 ii) dotted blue line, for $\xi=0.010$
iii) dashed red line, for $\xi=0.100$
iv) dot-dashed green line, for $\xi=0.150$.
}
\label{fig:entropy}
\end{figure*}

Now, we will move to the computation of the Ricci and the Kretschmann scalars. In this case, both quantities are slightly more complicated than the classical counterpart because the Schwarzschild ansatz is not valid. Firstly, the Ricci invariant, taking into account the two metric potentials, is then
\begin{align}
\begin{split}
    R = &-\frac{6 r^{12} \left(r^2-5 \xi ^2\right)}{\ell_0^2 (r-\xi )^7 (\xi +r)^7} 
    \\
    &
    -
    \frac{8 M_0 r^8 \left(\xi ^2+3 r^2\right) \left(\xi ^6-7 \xi ^2 r^4-12 \xi ^4 r^2+2 r^6\right)}{(r-\xi )^7 (\xi +r)^7 \left(\xi ^2+r^2\right)^2}
    \\
    &
    +\frac{48 M_0 r^{12} \left(r^2-5 \xi ^2\right) \ln \left(\frac{\xi ^2}{r^2}+1\right)}{\xi ^2 (r-\xi )^7 (\xi +r)^7} \,.
\end{split}
\end{align}
By taking just the dominant term in the $\xi$ expansion, we obtain
\begin{align}
    R &\approx R_0 \Bigg[ 
    1 + \left(\frac{8 \ell_0^2 M_0}{3 r^4} + \frac{2}{r^2}\right) \xi ^2
    \Bigg].
\end{align}
Secondly, we can compute the Kretschmann scalar $K$ defined as
\begin{align}
K \equiv R_{\alpha \beta \gamma \delta}R^{\alpha \beta \gamma \delta} .
\end{align}
Although possible, the complete expression is not simple, so we avoid showing it. Instead, we expand up to leading order in $\xi$, i.e., 
\begin{align}
    K \approx K_0 
    \Bigg[ 1 + 
   \left(
   \frac{16 \ell_0^2 M_0}{3 r^4}+\frac{4}{r^2}
   \right) 
   \xi ^2
    \Bigg].
\end{align}
We notice that the two scalars $\{R, K\}$ have singular points at $r \pm \xi=0$, which are just an artifact of the $\xi$ expansion competing with the limit of $r \rightarrow 0$.

\subsection{Thermodynamics} \label{sec:2}

In order to get a better comprehension of the underlying physics of this scale--dependent black hole solution, we investigate how the black hole thermodynamics is modified or if the inclusion of a running gravitation coupling modifies the physical features. We thus compute the Hawking temperature, the Bekenstein-Hawking entropy, and the specific heat.
First of all, the temperature is defined by
\begin{align} \label{temp}
T_H(r_H) &= \frac{1}{4 \pi} \Bigg|\lim_{r\rightarrow r_H} \frac{\partial_r g_{tt}}{\sqrt{-g_{tt}g_{rr}}} \Bigg|,
\end{align}
which yields
\begin{align} \label{temp_exact}
T_H(r_H) &= T_0(r_H)  \left[1 + \left(\frac{\xi}{r_H}\right)^{2}\right]^{-1}.
\end{align}
The above expression allows obtaining the black hole temperature in terms of the event horizon. An explicit form of the temperature is not possible to achieve, albeit we can expand when the SD parameter is close to zero to get,
\begin{align}\label{temp_approx}
T_H (\xi << 1) \approx T_0 (r_0) \Bigg[ 1 + \frac{1}{3} \left( \frac{\xi}{r_0}\right)^4  \Bigg].
\end{align}

Another useful window to understand the thermodynamic properties of a black hole is the Bekenstein-Hawking entropy.
It is well-known (from Brans-Dicke theory \cite{Jacobson:1993vj,Iyer:1995kg,Visser:1993nu,Creighton:1995au,Kang:1996rj}) that the entropy of black hole solutions in $D+1$ dimensions (with varying Newton's constant) satisfies the following relation:
\begin{align}\label{}
S_H(r_H)=\frac{1}{4}\oint_{r=r_H} \mathrm{d}^{D-1}x\frac{\sqrt{h}}{G(x)},
\end{align}
where $h_{ij}$ is the induced metric (at the horizon $r_H$). In 2+1 dimensions and circularly symmetric solution, the last integral is easy to compute. 
The induced metric for constant $t$ and $r$ slices is  $\mathrm{d}s=r\mathrm{d}\phi$. Also, $G(x)=G(r_H)$ is constant along the horizon. Finally, the entropy is
\begin{align}\label{Entropy}
S_H(r_H)&=\frac{2\pi r_H}{4G(r_H)} = S_0(r_H) \left[ 1 + \left(\frac{\xi}{r_H}\right)^2 \right].
\end{align}
Just like before, we cannot obtain analytical solutions for the entropy. 
Instead, an approximated solution is taken in the regime where $\xi$ is small. Thus, in terms of the classical black hole horizon, we have 
\begin{align}\label{Entropy2}
    S_H& (\xi << 1)\approx S_0(r_0) \Bigg[ 1 - \frac{1}{3} \left( \frac{\xi}{r_0} \right)^4  \Bigg].
\end{align}
Finally, regarding the specific heat, we consider the relation
\begin{align}
    C_H(r_H) &= S_H.
\end{align}
We thus confirm it is always positive. The later means that the positivity of the specific heat will ensure local thermal stability of the black hole \cite{Nadeem-ul-islam:2018mpc}. 

Of particular interest are re following comments:

\begin{enumerate}
    \item One observes that the inclusion of a running Newton's coupling makes the black hole horizon smaller than the classical counterpart. Such a feature profoundly impacts its thermodynamical properties, since relevant quantities are evaluated at the horizon.
    \item When $\xi$ is taken to be small, both the temperature and the entropy show corrections proportional to $\xi^4$. On the one hand, the temperature slightly increases with respect to the classical solution, and, on the other hand, the entropy slightly decreases, again, compared to the classical result.
    \item The pole located at $r \rightarrow \xi$ introduces additional features compared with the BTZ solution. 
    Thus, as can be observed in Fig.~\eqref{fig:horizons}, we only are considering vales  $r<\xi<\infty$. One further observes in Figs.~\eqref{fig:temperature} and \eqref{fig:entropy} that, 
    the classical solution and our solution converge
    for large $M_0$. In the opposite regime,
     for small values of the classical mass, we lose predictivity, and on-trivial features appear. Such effects strongly depend on the explicit form of Newton's coupling.
\end{enumerate}

\subsection{RG-improvement vs scale-dependence in black holes}

At this point, it is convenient to emphasize the conceptual differences between the most frequently used approach to incorporate quantum corrections in black hole physics, the RG-improvement formalism \cite{Bonanno:2000ep,Bonanno:2006eu,Koch:2013owa,Bonanno:2017zen,Pawlowski:2018swz}, and the methodology exploited in this work, the scale-dependence gravity \cite{Koch:2016uso,Rincon:2017ypd,Rincon:2017goj,Rincon:2017ayr,Contreras:2017eza,Rincon:2018sgd,Contreras:2018dhs,Rincon:2018lyd,Rincon:2018dsq,Contreras:2018gct,Canales:2018tbn,Rincon:2019cix,Panotopoulos:2021heb,Bargueno:2021nuc,Contreras:2018swc,Koch:2015nva,Contreras:2013hua,Alvarez:2020xmk,Rincon:2020cpz,Rincon:2020iwy}.
\newline
\begin{itemize}
    \item The traditional approach in RG-improvement formalism (introduced in \cite{Bonanno:2000ep}) demands that, after an adequate set of gravitational coupling, the classical black hole solutions are "adapted" through $G_0 \rightarrow G_k = G(k(r))$, and it is assumed that the classical background is properly "improved". The connection between the renormalization scale $k$ and the corresponding radial coordinate is always subject to speculation (for a detailed discussion, see \cite{Babic:2004ev,Koch:2014joa,Moti:2018rho} and references therein). However, one can infer an appropriate relation between these two parameters by dimensional analysis.\\
    This approach has the advantage that it explicitly incorporates the information from explicit quantum gravity calculations. Nevertheless, it has the disadvantage that the improved solutions are not solutions of 
    any known equations of motion or the extremum of a (known) effective action.
    \item 
    The scale-dependence gravity approach, in the form as it is used in \cite{Koch:2016uso,Rincon:2017ypd,Rincon:2017goj,Rincon:2017ayr,Contreras:2017eza,Rincon:2018sgd,Contreras:2018dhs,Rincon:2018lyd,Rincon:2018dsq,Contreras:2018gct,Canales:2018tbn,Rincon:2019cix,Panotopoulos:2021heb,Bargueno:2021nuc,Contreras:2018swc,Koch:2015nva,Contreras:2013hua,Alvarez:2020xmk,Rincon:2020cpz,Rincon:2020iwy} has the advantage that improved equations of motion (\ref{eq_eomSD}) are solved directly, making sure that any solution found for this is automatically extremizing a reasonable system. However, the shortcut of using energy conditions, came at the cost
    of loosing the information from the actual quantum gravity beta functions.
\end{itemize}
Thus, both approaches have advantages and disadvantages. In this paper, we used a mixture of both, taking only
the desirable features of each.
The information from the quantum gravity sector has been embedded through the scale-dependence
of the running gravitational coupling~(\ref{eq_Gr}), while the integration of the equation of motion~(\ref{eq_eomSD}) ensures consistent BH solutions. Therefore, the highlights of the RG-improvement and SD gravity have been incorporated in such that we obtain a well-defined black hole solutions in 2+1 dimensions.

\section{SD rotating solution}
In what follows we generalize the above solution and discussion to include rotation.
Let us start by considering maintaining rotational symmetry and considering a line element of the form
\be\label{lineelegen}
\mathrm{d}s^2= -f(r) \mathrm{d}t^2 + h(r) \mathrm{d}r^2 + r^2 \Bigl[N(r)\mathrm{d}t + \mathrm{d}\phi\Bigl]^2,
\ee
where $f(r)$, $h(r)$ $N(r)$ are functions that must be obtained from the effective Einstein's field equations. Also, following the same ideas of the non-rotating solution, we use the same concrete functional form of $G(r)$ and also obtain the cosmological function $\Lambda(r)$, algebraically.
Then all the other functions are obtained by analytically integrating the remaining equations of motion.
We obtain the set of functions $\{ f(r), h(r), \Lambda(r), G(r), N(r) \}$. 
In practice, this straight forward integration comes with several technical problems.  
The solutions can be consistently obtained as follows:
i) first, we compute de $G(r)$ using the same arguments as the non-rotating case
\begin{align}\label{eq_Gr2}
G(r) &= G_0 \Bigg[1 + \left(\frac{\xi}{r}\right)^2 \Bigg]^{-1}.
\end{align}
ii) We replace $G(r)$ into the effective equations, and also we eliminate $\Lambda(r)$ algebraically. Now, the task is to obtain $\{ f(r), h(r), N(r) \}$.
iii) We then compute $N(r)$, from the off-diagonal part of the field equations which reads
\begin{align}
\frac{\mathrm{d}}{\mathrm{d}r}\ln\Bigl(N'(r)\Bigl) &= \frac{7 \xi ^4 - 3 r^4 + 8 \xi ^2 r^2}{r \left(r^4-\xi ^4\right)}.
\end{align}
The last equation is a second order differential equation, which has two free parameters. They are selected in such a way that classical limit for the shift function, $N(r)$, is recovered. Thus, the solution take the simple form
\begin{align}
N(r) &= N_0(r) \delta(r,\xi),
\end{align}
where $N_0(r)$ corresponds to the classical shift function, i.e., 
\begin{align}
N_0(r) &= -\frac{J_0}{2r^2}
\end{align}
and the multiplicative correction $\delta(r,\xi)$ turns out to be
\begin{align}
\delta(r,\xi) &\equiv
\Bigg[
- 7
+ \frac{2 \xi ^2}{r^2} 
- \frac{\xi ^4}{3 r^4}
+ \frac{8 r^2}{\xi ^2}\ln \left(1 + \frac{\xi ^2}{r^2} \right)
\Bigg].
\end{align}
The integration constant is defined such that  the classical solution is recovered when $\xi \rightarrow 0$ on the function $\delta(r,\xi)$, namely
\begin{align}
&\lim_{\xi \rightarrow 0}\delta(r,\xi) = 1.
\end{align}
iv) Knowing $G(r)$ and $N(r)$, we can obtain a relation between $f(r)$ and $h(r)$ and verify that they are not reciprocal. Thus, we finally obtain, with help of the reduced equations, that
\begin{align} \label{hrotating}
h(r) &=  \Bigg[ 1 - \left(\frac{\xi}{r}\right)^2 \Bigg]^6   f(r)^{-1},
\end{align}
which coincides with the non-rotating case. 
v) Finally,  we will obtain the lapse function $f(r)$ from the remaining differential equation
\begin{align}
\frac{\mathrm{d}^2 f}{\mathrm{d}r^2 } + a(r)\frac{\mathrm{d} f}{\mathrm{d}r } + b(r)f + c(r) &= 0,
\end{align}
where the funcions $\{ a(r), b(r), c(r) \}$ are defined as
\begin{align}
a(r) &= -\frac{3 \xi ^4+r^4+8 \xi ^2 r^2}{r \left(r^4-\xi ^4\right)},
\\
b(r) &= \frac{8 \left(\xi ^4+2 \xi ^2 r^2\right)}{r^2 \left(r^4-\xi ^4\right)},
\\
c(r) &= -\frac{2 J_0^2 \left(r^2-\xi ^2\right)^7}{r^{12} \left(\xi ^2+r^2\right) \left(r^4-\xi ^4\right)}.
\end{align}
Again, the integration constants can be chosen in such a way that the generalized solution 
 mimics the classical black hole solution, namely
\begin{align}
f(r) &= -M(r) + \frac{r^2}{\ell_0^2} 
+ \frac{J(r)^2}{4r^2}.
\end{align}
At this point some commets are in order. Firstly, notice that the new solution resembles the classical one in its structure, although now the mass and angular momentum parameters are replaced by SD functions, which are defined as
\begin{align}
M(r) &= M_0 
\delta(r,\xi),
\\
J(r)^2 &= J_0^2 \delta(r,\xi)^2.
\end{align}
Second, when $J_0 \rightarrow 0$ the non-rotating case is obtained. Finally, notice that albeit $\Lambda_0 \rightarrow \Lambda(r)$, the term with such coupling parameter is not disturbed in the lapse function.  
vi) The function $h(r)$ is directly computed by using \eqref{hrotating}, to obtain
\begin{align}
h(r) &= \Bigg[ 1 - \left(\frac{\xi}{r}\right)^2 \Bigg]^6   
\Bigg[
-M(r) + \frac{r^2}{\ell_0^2} 
+ \frac{J(r)^2}{4r^2}
\Bigg]
^{-1}.
\end{align}
vii) As final step, we replace all the functions and their derivatives into the expression for the cosmological coupling $\Lambda(r)$ to finally obtain
\begin{align}
\begin{split}
\Lambda(r) &= \frac{1}{9 \ell_0^2 \xi ^4 \left(r^2-\xi ^2\right)^6 \left(\xi ^2+r^2\right)^2}
\\
& \Bigg[
24 \ell_0^2 r^8 \ln \left(1 + \frac{\xi ^2}{r^2} \right)
\Bigg(
2 J_0^2 
\Bigl(\xi ^8+6 \xi ^2 r^6 
\\
& - 15 \xi ^4 r^4-8 \xi ^6 r^2 \Bigl)
- 6 J_0^2 r^4 \left(r^2-3 \xi ^2\right) 
\\
& \times
\left(\xi ^2+r^2\right) \ln \left(1 + \frac{\xi ^2}{r^2} \right)
\\
& +3 M_0 \xi ^2 r^4 \left(r^2-3 \xi ^2\right) \left(\xi ^2+r^2\right)
\Bigg)
\\
& + \xi ^4 
	\Bigg(
 	-9 r^{12} \left(r^2-3 \xi ^2 \right) \left(\xi ^2+r^2\right)
 	\\
 	& -\left(-3 \xi ^6+12 r^6-39 \xi ^2 r^4+14 \xi ^4 r^2\right)
 	\\
 	& \times J_0^2 \ell_0^2 \left(-\xi ^6+12 r^6+3 \xi ^2 r^4+2 \xi ^4 r^2\right)
 	\\
 	& -12 \ell_0^2 M_0 r^8 \left(\xi ^6+6 r^6-15 \xi ^2 r^4-8 \xi ^4 r^2\right)
 	\Bigg) 
\Bigg].
\end{split}
\end{align}
The last function looks quite complicated, but we can check that it is the generalized cosmological coupling by simple substituting into the effective Einstein's field equations. As a sanity check, 
we can expand around $\xi \sim 0$.
 We find
\begin{align}
\Lambda(r) &= \Lambda_0 \left[ 1 + \frac{2\xi^2}{r^2} \right] + \mathcal{O}(\xi^3)
\end{align}
Now, with this rotating solution we will  briefly study the thermodynamics of the rotating case.
\subsection{Thermodynamics}
As we discuss in the non-rotating case, we can compute the Hawking temperature using the expression
\begin{align} \label{temp2}
T_H(r_H) &= \frac{1}{4 \pi} \Bigg|\lim_{r\rightarrow r_H} \frac{\partial_r g_{tt}}{\sqrt{-g_{tt}g_{rr}}} \Bigg|,
\end{align}
Thus, as the combination $\sqrt{-g_{tt}g_{rr}} \neq 1$, we expect to have a correction given by the SD Newton's coupling. Rewriting the expression we finally obtain
\begin{align}\label{THH}
T_H(r_H) &= \frac{1}{4 \pi}\Bigg|\frac{2 M_0 }{r_H} \left[1 + \left(\frac{\xi}{r_H}\right)^{2}\right]^{-1}  \Delta \Bigg|.
\end{align}
o more precisely
\begin{align}\label{THH}
T_H(r_H) &= \frac{1}{4 \pi}\Bigg|\frac{2 M_0 G(r_H) }{r_H} \Delta \Bigg|.
\end{align}
The last expression is the exact one for the black hole temperature. We observe that an SD Newton's coupling modifies the classical temperature, albeit the expression looks exactly the same that its classical counterpart replacing $G_0$ by $G(r)$. We also recognize that both the classical and SD solution temperature are vanishing for the extremal condition, i.e., when $J_0^{\text{max}} \equiv \ell_0 M_0$. 
In particular, we recall that $\Delta$ takes the usual form
\begin{align}\label{Delta}
\Delta &= \sqrt{1 -\Bigg( \frac{J_0}{M_0 \ell_0}\Bigg)^2}.
\end{align}
Finally, we will discuss the Bekenstein-Hawking entropy for the rotating case. 
In this case, the black hole horizon is then modified, but the definition of the entropy maintain its form, i.e., we can apply the same formula as before, but now considering the 
generalized (rotating) metric potentials. Thus, we have
\begin{align}\label{entropy_full}
S_H(r_H) &=\frac{\mathcal{A}_H(r_H)}{4G(r_H)} = S_0(r_H) \left[ 1 + \left(\frac{\xi}{r_H}\right)^2 \right].
\end{align}
We observe that both the temperature and the entropy generalize the classical solutions and are consistent with previous results in the context modified 2+1 dimensional black holes \cite{Rincon:2018lyd}.

\begin{figure*}[ht]
\centering
\includegraphics[width=0.48\textwidth]{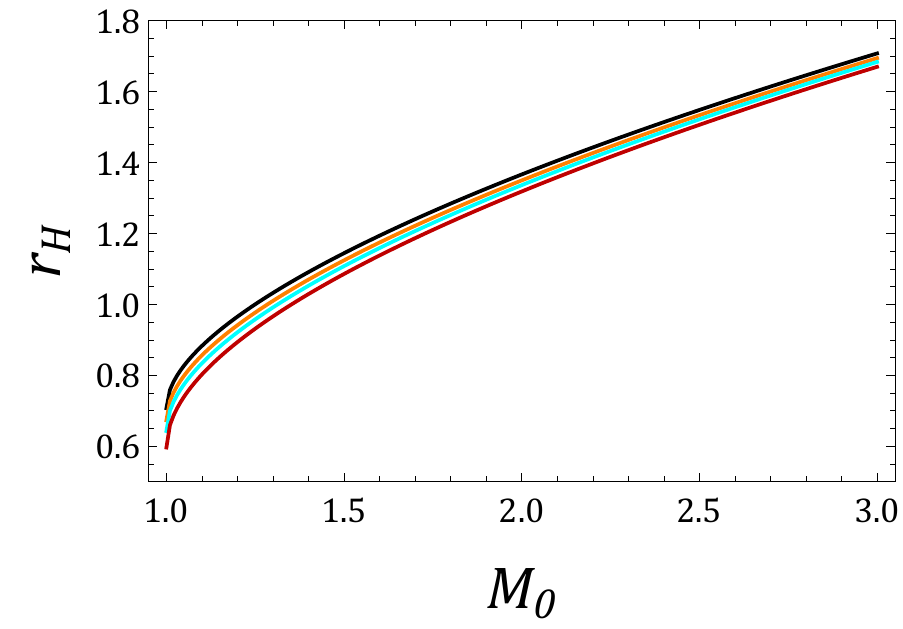}   
\ \ \
\includegraphics[width=0.48\textwidth]{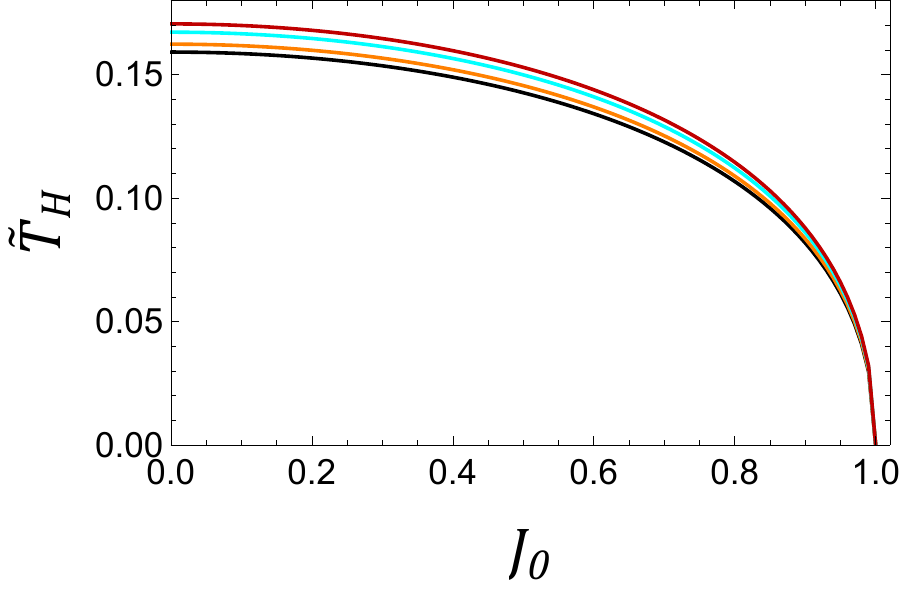}   
\caption{
{\bf{Left panel:}} Black hole horizon, $r_H$, against the classical black hole mass $M_0$,  for different values of the parameter $\xi$.
We have assumed $J_0 = 1, \ell_0 = 1$ and $G_0 = 1$. 
{\bf{Right panel:}} Scaled Black hole temperature, $\tilde{T}_H$, against the classical angular momentum parameter $J_0$ for different values of the parameter $\xi$. 
We have assumed $M_0 = 1, \ell_0 = 1$ and $G_0 = 1$. 
For educative purpuses we have rescale $T_H$ as follow: 
$\tilde{T}_H(\xi = 0.00) = 1.00 T_H$, 
$\tilde{T}_H(\xi = 0.15) = 1.02 T_H$, 
$\tilde{T}_H(\xi = 0.20) = 1.05 T_H$, 
$\tilde{T}_H(\xi = 0.25) = 1.07 T_H$, 
The color code is as follows:
  i) black line,  for $\xi=0.00$,
 ii) orange line, for $\xi=0.15$
iii) cyan line,   for $\xi=0.20$
 iv) red line,    for $\xi=0.25$.
 }
\label{fig:review}
\end{figure*}

Notice that both, temperature and entropy, depend on the black hole horizon, which is modified by the inclusion of angular momentum. Although $r_H$ is not analytical, we still can get insights by solving it numerically. Thus, we show in Fig. \eqref{fig:review} (left) the behaviour of the new horizon. We observe that, as occurs in the classical solution, the black hole horizon has a minimum value, determined by the condition $\Delta=0$, i.e., when the angular momentum takes its extreme value $J_0^{\text{max}} = M_0 \ell_0$. Thus, from that value the horizon increases. The Fig. \eqref{fig:review} (left) also shows that the scale-dependent black hole horizon is smaller than its classical counterpart, being consistent with other scale-dependent black hole solutions. Regarding the temperature,  $T_H$ has a natural cut-off which is reached when $\Delta=0$. 
For educative purposes, we plot $T_H$ against $J_0$ to show that, irrespectively of the value of $\xi$, such feature is preserved, namely,  a natural cut-off emerge for $J_0^{\text{max}} = M_0 \ell_0$, see Fig. \eqref{fig:review} (right). Also, the scale-dependent temperature is larger than its classical counterpart. To make the difference visible we exaggerate this effect in Fig. \eqref{fig:review} (right) rescaling properly the temperature (see labels).
%

\section{Concluding remarks} \label{conclu}
In the present article, we have studied effect a running gravitational coupling on the well--defined BTZ black hole solution offering an alternative view to previous works. 
A brief discussion regarding how the horizon is modified when Newton's coupling evolves as well as a comparison when the running parameter $\xi$ is taken as a small value is treated. 
Taking advantage of the lapse function, we have obtained the event horizon numerically, and we have contrasted its value with the approximated one. We observed that the new values of the event horizon are smaller than those obtained from the classical case, meaning that the black hole is strongly modified from a thermodynamic point of view.
Then, we computed the basic properties in black hole thermodynamics,i.e., the Hawking temperature, the Bekenstein Hawking entropy and specific heat. We observed that all the quantities are corrected by the incorporation of quantum features. We observe new poles appear on the invariants $R$ and $K$ absent in the classical solution. Of particular interest is the fact that the black hole is locally stable, verified via the computation of the specific heat, a result which is also true in the classical case.
We observe that when we consider Newton's coupling coming from AS as input, and include it in the SD formalism, the result is the violation of the NEC, 
and thus a violation of the Schwarzschild 
relation.
Last but not least, we briefly discussed the rotating case. 
We observe obtain an analytical solution wich is consistent 
 with the classical case \cite{Banados:1992wn} and with the original rotating SD version of the BTZ black hole  \cite{Rincon:2018lyd} in the appropriate limit(s). 
 We can obtain analytical expressions for shift function $N(r)$, the cosmological function $\Lambda(r)$, and the corresponding metric potentials $\{ f(r), h(r) \}$. 
We show that the thermodynamics is also robust, i.e., the temperature and the entropy maintain their functional form, in comparison with the classical counterpart. 
This reveals that, although non-trivial, the BTZ black hole is compatible with the inclusion of quantum features, making a minimum amount of changes and identifications.

\section*{Acknowledgments}

A.R. and N.C. acknowledge Universidad de Santiago de Chile for financial support through the Proyecto POSTDOCDICYT, C\'odigo 
042231CM-Postdoc.
The work of C.L. is funded by Becas Chile, ANID-PCHA/2020-72210073.


%
\bibliographystyle{unsrt}         
\bibliography{biblio_1}
%





\end{document}